\documentclass[aps,prb,twocolumn,fleqn]{revtex4}
\usepackage[latin1]{inputenc}
\usepackage{amsmath}
\usepackage{amsfonts}
\usepackage{amssymb}

\usepackage{graphicx} 
\def\gapx{\lower 2pt \hbox{$\buildrel>\over{\scriptstyle{\sim}}$\ }}
\def\lapx{\lower 2pt \hbox{$\buildrel<\over{\scriptstyle{\sim}}$\ }}
\def\ph2{{\it p}-H$_2$}
\def\beq{\begin{equation}}
\def\eeq{\end{equation}}
\def\Am3{\AA$^{-3}$}
\begin{document}

\widetext
\title{Pair potentials and  equation of state  of 
solid parahydrogen  to megabar pressure}
\author{Tokunbo Omiyinka and Massimo Boninsegni   } 
\affiliation{Department of Physics, University of Alberta, Edmonton, Alberta, Canada T6G 2G7}
\date{\today}

\begin{abstract}
We compute by means of Quantum  Monte Carlo simulations the equation of state of bulk solid {para}hydrogen extrapolated to zero temperature, up to  a pressure of $\sim$ 2 MBar. We compare the equation of state yielded by three different pair potentials, namely the Silvera-Goldman,  Buck and one recently proposed by Moraldi, modified at short distances to include a repulsive core, missing in the originally proposed potential. The Moraldi pair potential yields an equation of state in very good agreement with experiment at megabar pressures, owing to its softer core, and is at least as accurate as the SG or the Buck at saturated vapour pressure. Estimates for the experimentally measurable kinetic energy per molecule are provided for all pair potentials.
\end{abstract}
\maketitle

\section{Introduction}
Hydrogen is the simplest and most abundant element in the universe.  Achieving predictive knowledge of its equation of state in a broad range of thermodynamic conditions remains a worthwhile theoretical goal,  due to its relevance to a wide variety of physical systems, often with possible technological applications,\cite{mao,mao2,kohanoff,ashcroft} and also because it provides a cogent test of the most advanced computational many-body techniques.  \\ \indent 
The equation of state (EOS) of hydrogen, in essentially all  phases of interest, is a fully quantum-mechanical many-body problem. Theoretical calculations broadly fall into two different categories; in {\it ab initio} studies, the mathematical model explicitly takes into account the elementary constituents (namely electrons and protons), which interact via the electrostatic Coulomb potential. Calculations are carried out within the framework of Density Functional Theory (DFT)\cite{ward,barbee,hohl,scandolo,ogitsu,pickard} or Quantum Monte Carlo (QMC).\cite{alder,natoli,pierleoni,magro,holtzmann,delaney} On the other hand, in  molecular condensed phases one often adopts the Born-Oppenheimer approximation, allowing one to regard individual hydrogen molecules as elementary particles, and describe their interaction by means of a static potential, the simplest choice being that of a central pair potential. \\ \indent
An approach based on a pair potential only dependent on the distance between two molecules, obviously cannot describe any process involving electronic transfer, nor the energy contribution of interactions involving, e.g., triplets of molecules, nor any effect arising from the non-spherical character of the inter-molecular interaction; the importance of all of these physical mechanisms generally increases with the thermodynamic pressure of the condensed phase under investigation. However, there are  distinct advantages to the use of static pair-wise potentials, 
mainly that it is typically  computationally much faster and conceptually simpler than {\it ab initio} approaches. Furthermore, if pair potentials are used, thermodynamic properties of molecular hydrogen in the condensed phase can typically be computed essentially without any uncontrolled approximation, e.g., by means of QMC simulations.  
Thus, it is desirable to develop pair potentials affording a reasonably accurate, quantitative description of the condensed phase of hydrogen in broad ranges of thermodynamic conditions. 
\\ \indent 
The Silvera-Goldman model pair
potential\cite{SG} is arguably the most commonly adopted, and has been shown \cite{operetto} to afford a quantitative description of the  the low temperature, equilibrium solid phase of parahydrogen (the only species discussed here). Another popular pair potential is the Buck,\cite{buck1,buck2,colla} which is very similar to the Silvera-Goldman, featuring a slightly deeper attractive well. Both potentials yield an EOS in reasonable agreement with experiment at moderate pressure ($\lesssim$ 25 kPa). At higher pressure, however, they become increasingly inaccurate, leading to an overstimation of the pressure due to the excessive ``stiffness" of their repulsive core at 
short inter-molecular separation (below $\sim$ 2.8 \AA). 
This problem is also present in theoretical calculations of the EOS of condensed helium at high pressure, based on pair potentials; in that context, it is known that better agreement with the experimental EOS can be obained on including three-body terms, whose overall effect is that of softening the  repulsive core of the pair-wise interaction.\cite{bpc,pederiva,syc} Still, given the significant computational overhead involved in the inclusion of three-body terms, the question remains of whether a modified effective pair potential could offer more satisfactory agreement between theory and experiment at high (e.g., megabar) pressure.
\\ \indent
To this aim, Moraldi recently proposed\cite{moraldi} a modified effective pair potential, with a softened repulsive core, designed to afford greater agreement between the theoretically predicted values of the pressure as a function of the density in the limit of low temperature ($T\to 0$). Results of perturbative calculations carried out in Ref. \onlinecite{moraldi} show remarkable agreement with experimental data of the pressure computed with the modified pair potential, in an extended range of density (up to 0.24   \AA$^{-3}$, which corresponds to a pressure in the megabar range).
\\ \indent
In this work, the $T=0$ EOS of  solid parahydrogen is computed by means of first principle QMC simulations, in a wider density range with respect to Ref. \onlinecite {moraldi}, namely up to a density of 0.273 \AA$^{-3}$, for a model of condensed parahydrogen based on three different pair potentials, namely a slightly modified version of  the Moraldi potential (see below), the Silvera-Goldmann and the Buck.  Our purpose is that of providing a numerically accurate, non-perturbative check of the pressure computed in Ref. \onlinecite{moraldi}, and to offer a cogent comparison of the performance of the three potentials. We also compute the total and kinetic energy per molecule; the latter is experimentally measurable by means of neutron scattering, and is especially sensitive to the detailed features of the repulsive core of the interaction.\cite{bpc}
\\ \indent 
We find that the recently proposed pair potential indeed yields pressure estimates in much better agreement with experiment than the Silvera-Goldman and Buck potential, especially above 0.14 \AA$^{-3}$. In fact, the agreement between the values of the pressure computed with the modified Moraldi potential and the experimentally measured ones remains excellent up to $\sim$ 180 GPa, the highest pressure for which experimental data are available for solid H$_2$. At the equilibrium density, i.e., zero pressure at $T$=0, the Moraldi potential yields energy estimates comparable to those of the Silvera-Goldman. Kinetic energy estimates at high density computed with the Moraldi potential are in significant disaccord with (considerably lower than) those furnished by the Silvera-Goldman and Buck potentials, a fact ascribable to the softer repulsive core of the Moraldi pair interaction. A measurement of the kinetic energy will therefore furnish additional important insight on whether the  Moraldi potential not only yields an EOS in quantitatively close agreement with experiment, but also affords a generally more accurate, quantitatively physical description of the system than the Silvera-Goldman or Buck model interactions.
\\
\indent
 In the next section, the microscopic model underlying the calculation, specifically the pair potentials utilized are illustrated; we briefly outline the basic methodology utilized and offer relevant computational details in Sec. \ref{comp}. In Sec. \ref{Results} a thorough illustration of the results obtained in this work is provided; finally, a general discussion is offered, and conclusions outlined, in Sec. \ref{Disc}.
\section{Model}
\begin{figure}[h]
\centerline{\includegraphics[height=2.4in,angle=0]{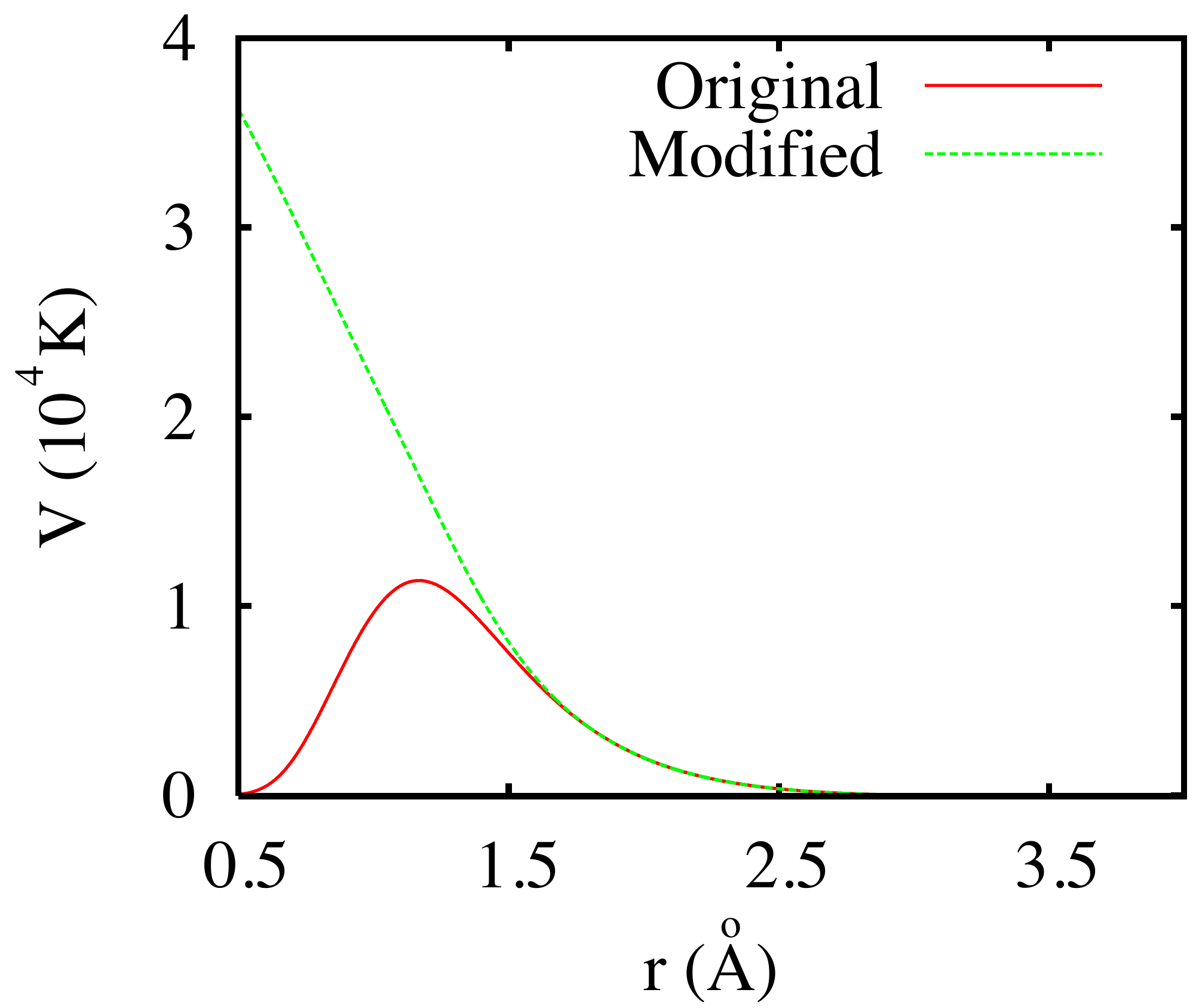}}
\caption{{\it Color online}. The Moraldi pair potential  $V_M(r)$ as proposed in Ref. \onlinecite{moraldi} (solid line) and the modified one $U(r)$ 
used in the present work (dashed line).}\label{f1}
\end{figure}

Our system of interest
is modeled as an ensemble of $N$ parahydrogen molecules, regarded as point particles of spin zero,
enclosed in a vessel of volume $\Omega$, shaped as a parallelepiped, with periodic boundary conditions in all directions. The sides of the parallelepiped are chosen to fit a crystalline sample of solid parahydrogen (hexagonal close-packed structure).
The quantum-mechanical many-body Hamiltonian is the following:
\begin{equation}\label{one}
\hat H = -\lambda\sum_{i=1}^N \nabla_i^2 + \sum_{i<j} V(r_{ij}) 
\end{equation}
Here, $\lambda$=12.031 K\AA$^2$, while  
$V$ is the potential describing the interaction between two molecules, 
only depending on their relative distance. Because this work is specifically about comparing different potentials, we provide here some basic 
details  of all potentials used. The Silvera-Goldman ($V_{SG})$ and Buck ($V_B$) potentials have the form
\beq
V_{}(r)=V_{rep}(r)-V_{att}(r)f_C(r)
\eeq
where $V_{rep}(r)$ describes the repulsion of two molecules at short distances, arising mainly from Pauli exclusion principle, preventing the electronic clouds of different molecules from spatially overlapping, whereas $V_{att}(r)$ represents the long-range Van der Waals attraction of mutually induced molecular electric dipoles. The function $f_C(r)$ has the purpose of removing the divergence of $V_{att}(r)$ as $r\to 0$. The detailed forms of these functions, as well as the values of the parameters for the two potentials, are furnished in Refs. \onlinecite{SG} and \onlinecite {buck1}.\\ \indent
The Moraldi potential, as proposed in Ref. \onlinecite{moraldi}, is expressed as follows:
\beq
{V_M}(r) = V_{rep}(r)f_{rep}(r)-V_{att}(r)f_C(r)
\eeq
where the functions $V_{rep}$, $V_{att}$ and $f_C$ are those of the Silvera-Goldman potential, whereas $f_{rep}(r)$ is a function of the same form of $f_C$, softening the repulsion at short distances contained in $V_{rep}$. As shown in Ref. \onlinecite{moraldi}, the potential $V_M(r)$ displays a much slower growth as $r\to 0$ than the Silvera-Goldman. In fact, the function $V_M(r)$ stops growing altogether at $r\sim r_\circ=1.28$ \AA, and actually {\it decreases} all the way to zero as $r\to 0$ (see Figure \ref{f1}).
\\
This  unphysical feature of $V_M(r)$  is of little consequences in calculations at low density, but must be corrected in order to prevent particles from ``tunnelling" across the potential barrier and falling on top of one another, something observed in computer simulations making use of $V_M(r)$ at densities for which the inter-molecular distance is  $\lesssim$ 2 \AA. Thus, in this work we made use of the modified  potential $U(r)$, defined as follows:
\begin{eqnarray}
U(r)=V_M(r)\Theta(r-d)+{\tilde V}_{rep}(r)\Theta(d-r)
\end{eqnarray}
where $d$=4 a.u. (i.e., 2.1167 \AA), $\Theta(x)$ is the Heaviside function and ${\tilde V}_{rep}(r)$ has the same form of $V_{rep}$, but the
two parameters (usually referred to as $\alpha$ and $\beta$) upon which it depends are adjusted to match $V_M(r)$ and its first derivative at $r=d$. This modified potential, utilized in our simulations, is shown alongside $V_M(r)$ in Figure \ref{f1}. It does not have the unphysical feature of $V_M(r)$ at short distances discussed above, indeed it increases monotonically as $r\to 0$.
\\ \indent
The above modification of the Moraldi potential is obviously one of the many that are possible; it has the advantage of being easy to implement. The detailed form of the repulsive core of the potential at molecular separation shorter than $\sim$ 1.5 \AA\ turns out not to affect the calculation significantly, in that particles seldom find themselves at such distances, in the range of pressure explored here. In other words, a repulsive core is needed, in order to avoid the unphysical behavior described above, but its precise form is not crucial, in the range of density considered in this work.

\section{Calculation}\label{comp}
The $T=0$ (ground state) equation of state of solid parahydrogen, modelled by the many-body Hamiltonian (\ref{one}) with the three different potentials $V_{SG}$, $V_B$ and $U$ described above, was computed by means of numerical QMC simulations based on the continuous-space Worm 
Algorithm (WA).\cite{worm,worm2,cuervo} Specifically, we estimate the pressure $P$ as a function of the density $\rho$, in the interval 
$0.0763$ \AA$^{-3} \le \rho\le 0.273$ \AA$^{-3}$, for a finite system comprising $N$=216 molecules. In order to provide ground state estimates, we perform simulations at different low temperatures and extrapolate the results to $T=0$. We generally find that results obtained at $T=4$ K are indistinguishable from the extrapolated ones, within our quoted statistical uncertainties. We use the standard virial estimator\cite{imada} for the pressure, and estimate the contribution from particles lying outside the main simulation cell by radially integrating the quantity $r (dV/dr)$, 
approximating the pair correlation function $g(r)$ to 1 (we do the same for the potential energy).\cite{correction}
The QMC results presented here are  obtained using a time step $\tau=5\times10^{-4}$ K$^{-1}$, as  we observed that the estimates yielded by such a choice of time step coincide with those extrapolated to $\tau=0$, within statistical uncertainties.
\\ \indent
It should be mentioned that in principle parahydrogen molecules must be regarded as indistinguishable particles of spin zero, and thus obeying Bose statistics. The WA explicitly allows for permutations of identical particles, and quantum-mechanical exchanges can be important in parahydrogen, in specific circumstances. For example, they are predicted to underlie superfluidity at low temperature in small clusters (less than thirty molecules),\cite{fabio,fabio2} and their effect is measurable in the momentum distribution of the liquid near melting.\cite{me}  In the solid phase, however, 
in the range of density explored here, exchanges are practically absent, i.e., particles can be regarded for practical purposes as distinguishable.

\section{Results}\label{Results}

Although the focus of our work is directed to the ground state EOS of solid parahydrogen at relatively high (megabar) pressure, we begin the discussion of the results by illustrating results of QMC simulations at the saturation density  $\rho_\circ=0.0261$ \AA$^{-3}$. Specifically, we compare the pressure and energy estimates yielded by the 
Moraldi potential (modified as explained above), which was not purposefully designed to describe the low-density equilibrium phase,   
to those yielded by the Silvera-Goldman and Buck potentials, as well as to experimental data. The purpose of such a comparison is that of ascertaining whether the softening of the Silvera-Goldman repulsive core, which is at the basis of the Moraldi potential,  might worsen the agreement with experiment at low density, while possibly improving it in the megabar pressure range.
\begin{table}[h]
\begin{tabular}{ccc} \hline\hline
&This work&Ref. \onlinecite{operetto}  \\ \hline
Moraldi  &-88.5(2) &  \\
Silvera-Goldman  &-88.1(1) &-87.90(2)  \\
Buck  &-93.8(1) &-93.87(2) \\ \hline\hline 
\end{tabular}\hfil\break 
\caption {\noindent {Ground state energy per molecule (in K) for solid parahydrogen in the {\it hcp} phase at the saturation density $\rho_\circ =0.0261$ \AA$^{-3}$, computed in this work and in
Ref. \onlinecite{operetto} for various intermolecular potentials. Statistical errors, in parentheses, are on the last digit.}
}\label{tableone}
\end{table} 

Table \ref{tableone} shows numerical estimates of the ground state energy per molecule for solid parahydrogen at saturation density. The estimates obtained here, obtained by extrapolating all the way down to $T=0$ results obtained in the range 1 K $\le T \le $ 4 K, are in agreement with those obtained in Ref. \onlinecite{operetto} using ground state 
Diffusion Monte Carlo (DMC) simulations. Experimentally, the ground state energy per molecule has been inferred in different ways, and the agreement between the various estimates\cite{simon,stewart,schnepp,driessen}  is not perfect; quoted values range from $-89.9$ K of Ref. \onlinecite{schnepp} to $-93.5$ K of Refs. \onlinecite{stewart} and \onlinecite{driessen}. It is clear from the data reported in the table that the Moraldi potential affords an accuracy comparable to that of the other two potentials.
\begin{figure}[h]
\centerline{\includegraphics[height=2.4in, angle=0]{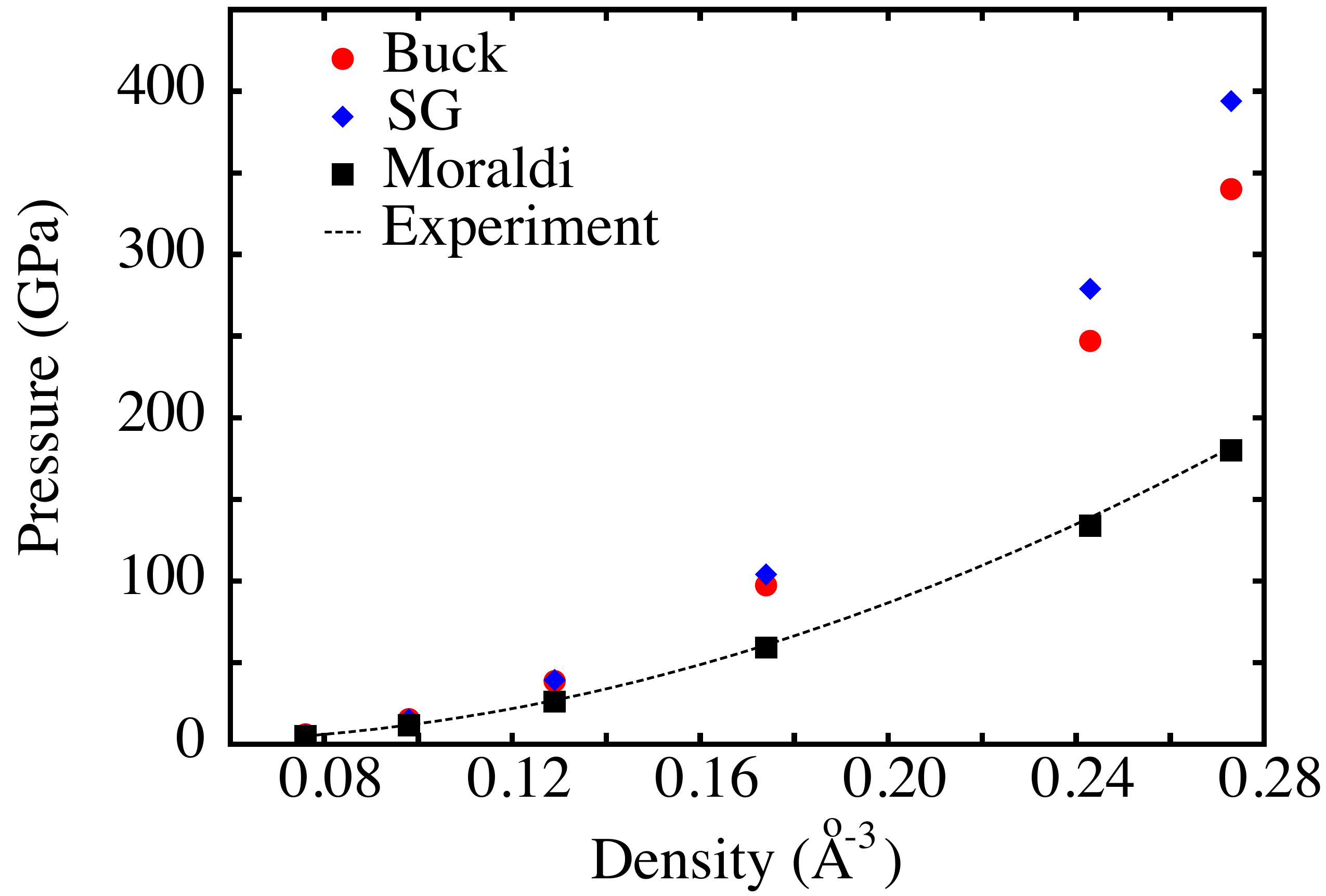}}
\caption{{\it Color online}. Equation of state of solid parahydrogen at $T$=0, computed by Quantum Monte Carlo simulations using the Silvera-Goldman (diamonds), Buck (circles) and Moraldi (boxes) pair potentials. Dashed line represents a fit to experimental data from Ref. \onlinecite{Freiman} obtained with the Vinet equation of state (see text). Combined statistical and systematic errors affecting our computed pressures are at the most of the order of 0.1\% of the pressure. }\label{comparison}
\end{figure}

We compare in figure \ref{comparison} our computed values of the pressure for {\it hcp} parahydrogen in the $T\to 0$ limit, in a density range  up to 0.273 \AA$^{-3}$, which corresponds to a pressure of about 1.8 Mbars. We also include for comparison a fit to the most recent\cite{Freiman} compilation of experimental (x-ray) measurements\cite{mao,loubeyre,aka} of the EOS obtained with the Vinet equation of state,\cite{vinet}
\beq
P =\frac{3K_\circ(1-d)}{d^2} \ {\rm exp}\biggl [\frac{3}{2}(K_\circ^\prime-1)(1-d)\biggr ]
\eeq
 where $d\equiv (\rho_\circ/\rho)^{1/3}$, with parameters $K_\circ=0.20(1)$ GPa and $K_\circ^\prime=6.84(7)$.
The computed values of the pressure are also reported in Table \ref{tabletwo}.
\begin{table}[h]
\begin{tabular}{cccc} \hline\hline
Density &Moraldi&Buck&SG\\ \hline
0.076&$5.11\times 10^4$ &$6.09\times 10^4$ &$5.92\times 10^4$ \\ 
0.098&$1.17\times 10^5$&$1.54\times 10^5$ & $1.52\times 10^5$\\
0.129&$2.61\times 10^5$&$3.87\times 10^5$ & $3.93\times 10^5$ \\
0.174&$5.93\times 10^5$&$9.74\times 10^5$ &$1.04\times 10^6$\\
0.243&$1.34\times 10^6$&$2.47\times 10^6$ &$2.79\times 10^6$ \\
0.273&$1.80\times 10^6$&$3.40\times 10^6$ &$3.94\times 10^6$ \\ \hline
\hline
\end{tabular}
\hfil\break 
\caption {\noindent {Pressure (in bars) calculated for different densities (in \AA$^{-3}$) in the $T\to 0$ limit, using Buck, Silvera-Goldman and Moraldi potentials. Combined statistical and systematic errors affecting our QMC simulations are estimated to be of the order of 0.1\% or less of the 
quoted values.}
}\label{tabletwo}
\end{table} 

\begin{figure}[h]
\centerline{\includegraphics[height=2.4in,angle=0]{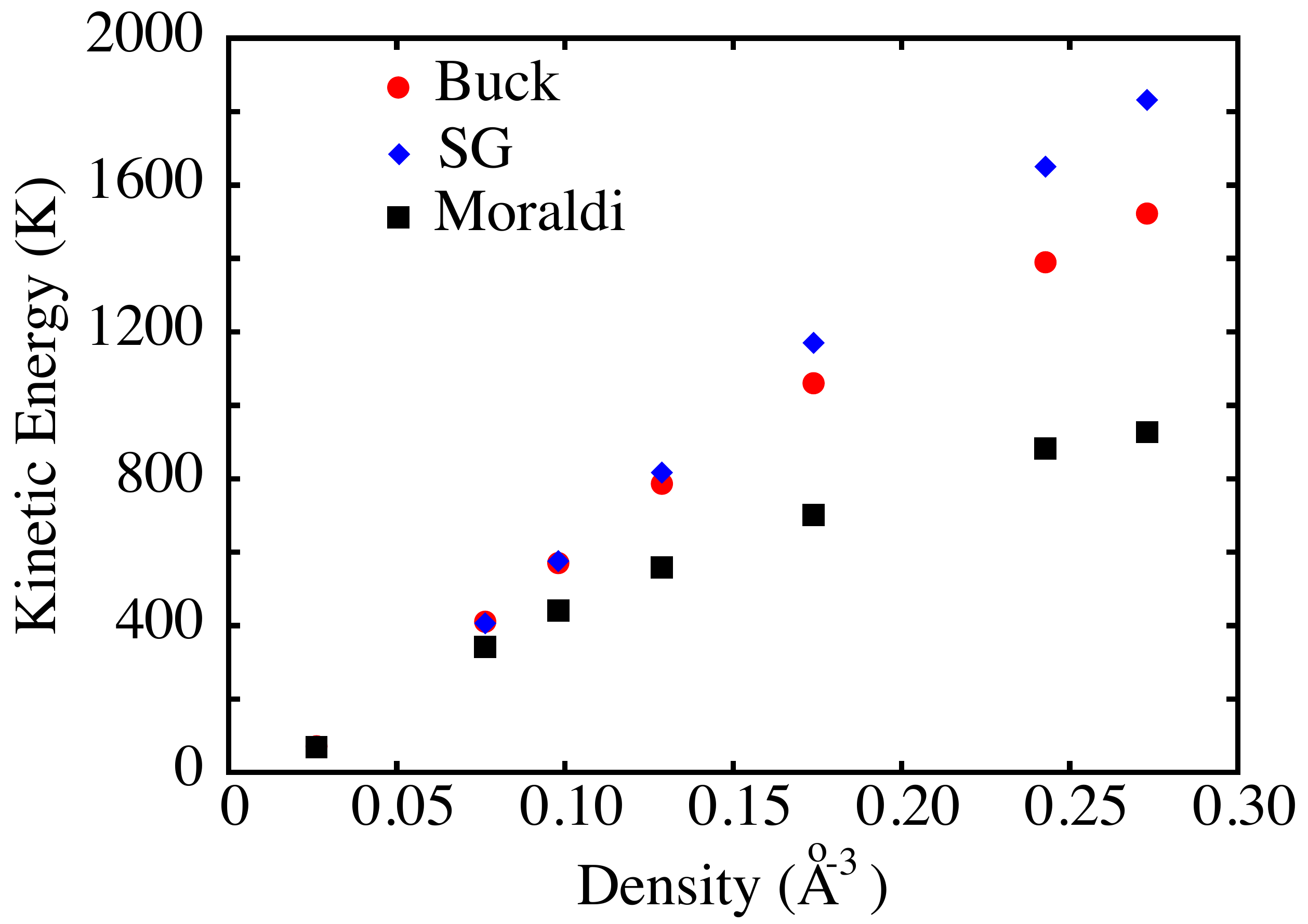}}
\caption{{\it Color online}. Kinetic energies per molecule at $T$=0, computed by simulations at different densities for the three potentials considered here}\label{ke}
\end{figure}

The agreement between the experimental values of the pressure, and those obtained by simulation using the Moraldi potential is excellent, considering the relatively simple form of the pair potential itself. At the highest compression considered here, namely 0.273 \AA$^{-3}$, the Silvera-Goldman potential overestimate the pressure by over a factor two. The Buck potential does only marginally better than the Silvera-Goldman. The lower values of the pressure yielded by the Moraldi potential are a direct consequence of its softer repulsive core. It should be noted that the values in Table \ref{tabletwo} are in quantitative agreement with those obtained by Moraldi in Ref. \onlinecite{moraldi},  based on a perturbative calculation. Perhaps even more significant is the fact that agreement remains excellent even above 110 GPa, i.e., a compression at which solid hydrogen undergoes an orientational phase transition.\cite{lorenzana} One would expect that in the orientationally ordered phase, the approximation of a spherically symmetric potential would be particularly problematic. 
\begin{figure}
\centering
\begin{tabular}{cc}
\includegraphics[height=0.36\columnwidth]{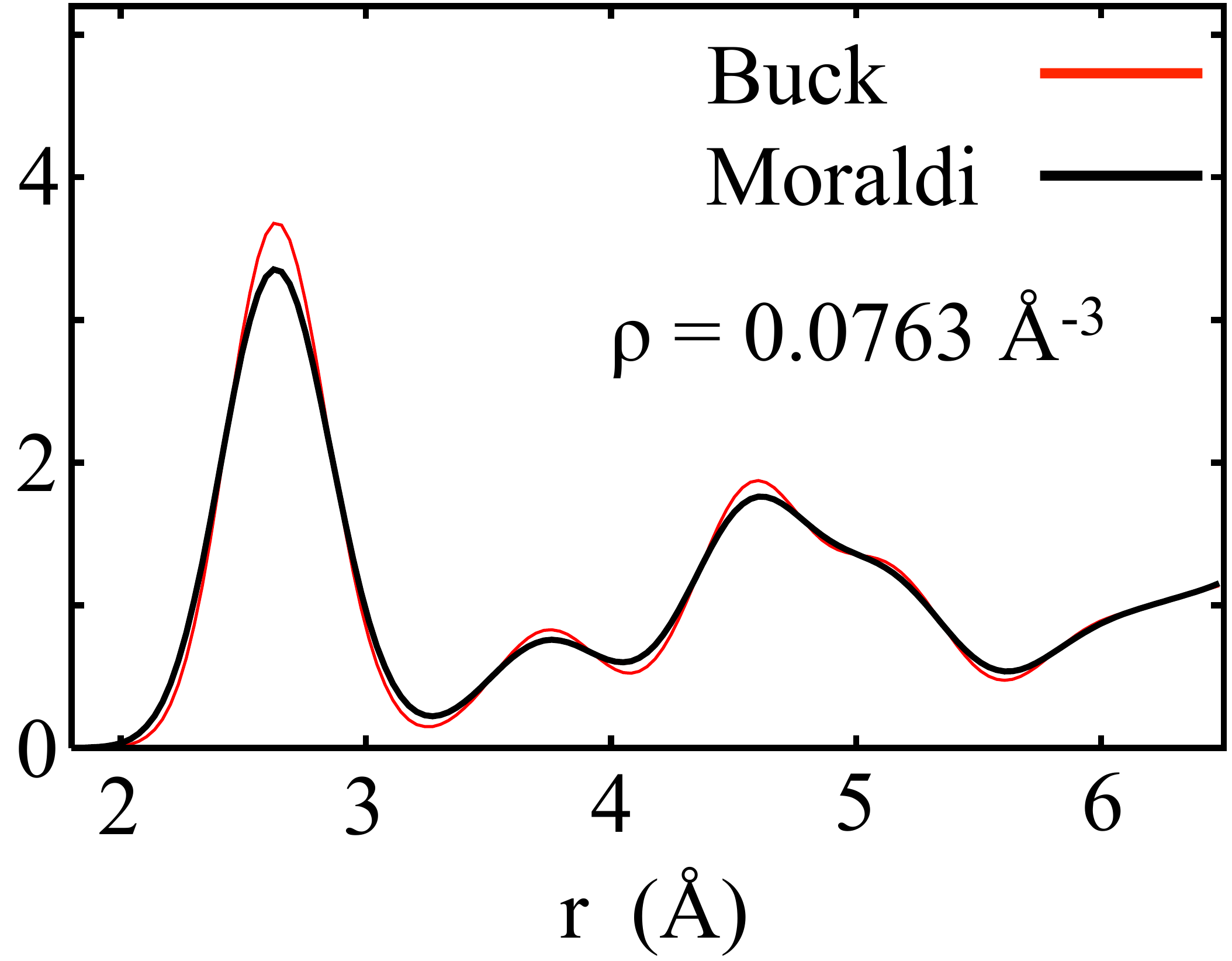}&
\includegraphics[height=0.36\columnwidth]{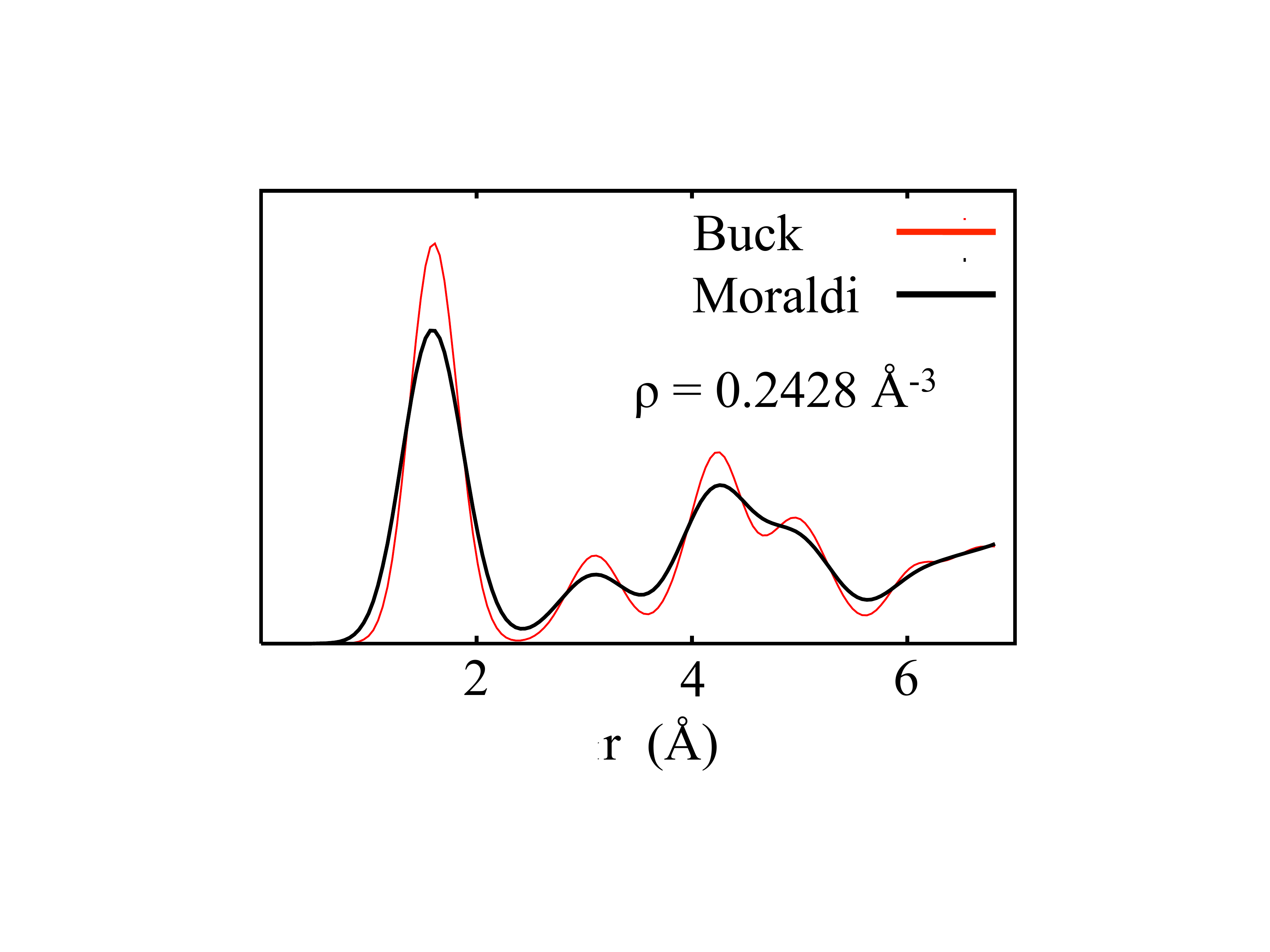}
\end{tabular}
\caption{{\it Color online}. Pair correlation function $g(r)$ for hcp parahydrogen at $T$=4 K, computed by simulation  at the two densities $\rho$=0.0763 \AA$^{-3}$ and $\rho$=0.2428 \AA$^{-3}$, with the Moraldi and Buck potentials. The results for the Silvera-Goldman potential are similar to those for the Buck.}\label{f4p}
\end{figure}
\\ \indent
\indent
The fact that the Moraldi potential is not as stiff as the Sivera-Goldman or Buck potentials at short distances is also reflected on the predicted structural properties of the crystal. In particular, the local environment experienced by a single atom is captured by the static structure factor,  which of course can be probed by X-ray diffraction. Its Fourier transform, the pair correlation function $g(r)$, is easily accessible by simulation. 
Results for two different densities are shown in Fig. \ref{f4p}, where the pair correlation functions computed for the Buck and Moraldi potentials are compared. It should be noted that the SG and Buck potentials yield similar results for this quantity (the main peak of the $g(r)$ is $\sim$ 10\% higher with the SG potential, at the higher density). While at the lower density, which corresponds to a pressure around 5 kilobars, the pair correlation functions yielded by the two potentials are comparable, at the higher density  (which corresponds to a pressure close to one Megabar), there is a clear difference between the two results. Specifically, the pair correlation function computed with the Buck potential displays considerably higher peaks and generally much sharper features and an overall more classical structure, compared to that obtained with the softer Moraldi potential, which is smoother, and  displays a markedly more quantum-mechanical character. 
\\ \indent 
While the improvement afforded by the Moraldi potential on the calculation of the EOS is certainly remarkable, especially considering the relatively wide range of pressure for which it is observed, some care should be exercised if a comprehensive assessment is sought of the  microscopic physical description yielded by a given pair potential. It is possible to obtain, by means of an effective pair potential, an EOS in close agreement with experiment, but at the cost of worsening the agreement for other quantities. \\
\indent
In particular, the kinetic energy per particle is particularly sensitive to the detailed shape of the repulsive core of the intermolecular potential at short distances (see, for instance, discussion in Ref. \onlinecite{bpc}), and it can also be measured experimentally by neutron scattering, as the second moment of the single-particle momentum distribution.\cite{glyde} Thus, a cogent test of the relative accuracy of the new potential with respect to the existing ones could be offered by measurements of the kinetic energy per particle, for which we have obtained ground state estimates as well, for all three potentials considered here. The results are displayed in Fig. \ref{ke} and detailed in Table \ref{tablethree}.
\begin{table}[h]
\begin{tabular}{ccccc} \hline\hline
Density&Moraldi&SG&Buck&Ref.\onlinecite{moraldi}\\ \hline
0.0261&69&70&71& \\
0.0763&341&406&411&405\\
0.0980&440&576&570&531\\
0.1288&558&817&786&678\\
0.1739&701&1170&1060&834\\
0.2428&872&1650&1390&914\\ 
0.2730&927&1832& 1523&  \\ \hline\hline
\end{tabular}
\hfil\break 
\caption {\noindent {Kinetic energy per molecule (in K) in the $T\to 0$ limit, at various densities (in \AA$^{-3}$), obtained  using each of the 
three pair potentials considered in this work in the QMC simulation. Combined statistical and systematic errors affecting our results are estimated at one percent  or less of the quoted values. Also reported are the estimates from Ref. \onlinecite{moraldi}, in the rightmost column.}
}\label{tablethree}
\end{table} 

At the saturation density $\rho_\circ$ all potentials yield essentially the same kinetic energy per molecule, whereas at higher density the softer core of the Moraldi potential results in a considerably lower kinetic energy, especially at the highest compression. It should be noted, in any case, that the kinetic energy contribution to the pressure is negligible compared to that arising from intermolecular interactions.  A comparison of the estimates obtained in this work with those reported in Ref. \onlinecite{moraldi} shows that the perturbative approach leads to a slight overestimation, but otherwise furnishes reasonably accurate results for the problem of interest.

\section{Conclusions}\label{Disc}
In this work we have carried out a numerical analysis of a new intermolecular potential proposed by Moraldi, aimed at reproducing the experimental equation of state of solid parahydrgen at low temperature, up to a pressure of 1.8 Mbars. The potential considered here is a  modification of the one proposed in Ref. \onlinecite {moraldi}, to which a repulsive core has been added at short distances, to prevent unphysical behavior at the highest densities considered. This potential has a considerably softer core than the Silvera-Goldman and Buck ptential, resulting in enhanced quantum effects and lower values of the pressure. We have carried out Quantum Monte Carlo simulation of {\it hcp} parahydrogen in the $T\to 0$ limit, and compared estimates yielded by the new potential, as well 
as the Silvera-Goldman and Buck potentials, for the pressure and for the kinetic energy per molecule, as a function of density. 
We have also compared the numerical results to the most recent experimental measurements of the equation of state, and found that the new potential yields excellent agreement with experiment, agreement which extends all the way down to low density (saturation).
We  also furnish estimates for the kinetic energy per molecule, whose measurement by neutron diffraction will provide another independent, important assessment of the reliability of the new potential.
\\
\section*{Acknowledgments}
The authors wish to acknowledge useful discussions with M. Moraldi.


\begin{thebibliography}{03}
\bibitem{mao}
H.-K. Mao and R. J. Hemley, Rev. Mod. Phys. {\bf 66}, 671 (1994).
\bibitem{mao2}
R. J. Hemley and H.-K. Mao, J. Low Temp. Phys. {\bf 122}, 331 (2001).
\bibitem{kohanoff}
J. Kohanoff, J. Low Temp. Phys. {\bf 122}, 297 (2001), and references therein.
\bibitem{ashcroft}
N. W. Ashcroft in {\it Proceedings of the International School of Physics ``Enrico Fermi," Course CXLVII, High pressure Phenomena}, edited by R. J. Hemley, G. L. Chiarotti, M. Bernasconi, and L. Ulivi (IOS, Amsterdam, 2002), p. 151.
\bibitem{ward}
M. W. C. Dharma-wardana  and F. Perrot, Phys. Rev. A {\bf 26}, 2096 (1982).
\bibitem{barbee}
T. W. Barbee, M. L. Coehn and J. L. Martins, Phys. Rev. Lett. {\bf 62}, 1150 (1989).
\bibitem{hohl}
D. Hohl, V. Natoli, D. M. Ceperley and R. M. Martin, Phys. Rev. Lett. {\bf 71}, 541 (1993).
\bibitem{scandolo}
J. Kohanoff, S.  Scandolo, G. L. Chiarotti and E. Tosatti,
Phys. Rev. Lett. {\bf 78}, 2783 (1997). 
\bibitem{ogitsu}
H. Kitamura, S. Tsuneyuki, T. Ogitsu and T. Miyake, Nature {\bf 404}, 259 (2000).
\bibitem{pickard}
C. J. Pickard and R. J. Needs, Nature Physics {\bf 3}, 473 (2007).
\bibitem{alder}
D. M. Ceperley and B. J. Alder, Phys. Rev. B {\bf 36}, 2092 (1987).
\bibitem{natoli}
V. Natoli, R. M. Martin and D. M. Ceperley, Phys. Rev. Lett. {\bf 70}, 1952 (1993).
\bibitem{pierleoni}
C. Pierleoni, D. M. Ceperley, B. Bernu, W. R. Magro, Phys. Rev. Lett. {\bf 73}, 2145 (1994).
\bibitem{magro}
W. R. Magro, D. M. Ceperley, C. Pierleoni and B. Bernu, Phys. Rev. Lett. {\bf 76}, 1240 (1996).
\bibitem{holtzmann}
C. Pierleoni, D. M.  Ceperley and M. Holzmann, Phys. Rev. Lett. {\bf 93}, 146402 (2004).
\bibitem{delaney}
K. T.  Delaney, C. Pierleoni and D. M. Ceperley, Phys. Rev. Lett. {\bf 97}, 235702 (2006).
\bibitem{SG}
I. F. Silvera and V. V. Goldman, J. Chem. Phys. {\bf 69}, 4209 (1978).
\bibitem{operetto}
F. Operetto and F. Pederiva,  Phys. Rev. B {\bf 73}, 184124 (2006).
\bibitem{buck1} U. Buck, F. Huisken, A. Kohlhase, D. Otten, and J. Schaeffer,
J. Chem. Phys. {\bf 78}, 4439 (1983).
\bibitem{buck2} M. J. Norman, R. O. Watts and U. Buck, J. Chem. Phys.
{\bf 81}, 3500 (1984). 
\bibitem{colla}
For a comparison of the Silvera-Goldman and Buck potentials,
see, for instance, E. Cheng and K. B. Whaley, J.
Chem. Phys. {\bf 104}, 3155 (1996).
\bibitem{bpc}
M. Boninsegni, C. Pierleoni and D. M. Ceperley, Phys. Rev. Lett. {\bf 72}, 1854 (1994).
\bibitem{pederiva}
S. Moroni, F. Pederiva, S. Fantoni and M. Boninsegni, Phys. Rev. Lett. {\bf 84}, 2650 (2000).
\bibitem {syc}
S.-Y. Chang and M. Boninsegni, J. Chem. Phys. {\bf 115}, 2629 (2001).
\bibitem{moraldi}
M. Moraldi, 
J. Low Temp. Phys. {\bf 168}, 275 (2012).
\bibitem{worm}
M. Boninsegni, N. V. Prokof'ev and B. V. Svistunov, Phys. Rev. Lett. {\bf 96}, 070601 (2006).
\bibitem{worm2}
M. Boninsegni, N. V. Prokof'ev and B. V. Svistunov, Phys. Rev. E {\bf 74}, 036701 (2006).
\bibitem{cuervo}
J. E. Cuervo, P.-N. Roy and M. Boninsegni, J. Chem. Phys. {\bf 122}, 114504 (2005).
\bibitem{imada}
M. Takahashi and M. Imada, J. Phys. Soc. Jpn. {\bf 53}, 3871 (1984).
\bibitem{correction}
At the highest density considered here, the estimated correction to the computed pressure arising from the tail of the intermolecular potential has a magnitude worth approximately 3\% of the total pressure.
\bibitem{fabio}
F. Mezzacapo and M. Boninsegni, Phys. Rev. Lett. {\bf 97}, 045301 (2006).
\bibitem{fabio2}
F. Mezzacapo and M. Boninsegni, Phys. Rev. A {\bf 75}, 033201 (2007).
\bibitem{me}
M. Boninsegni, Phys. Rev. B {\bf 79}, 174203 (2009).
\bibitem{simon}
F. Simon, Z. Phys. {\bf 15}, 307 (1923).
\bibitem{stewart}
J. Stewart, J. Chem. Phys. Solids {\bf 1}, 146 (1956).
\bibitem{schnepp}
O. Schnepp, Phys. Rev. A {\bf 2}, 2574 (1970).
\bibitem{driessen}
A. Driessen, J. A. de Waal and I. F. Silvera, J. Low Temp. Phys. {\bf 34}, 255 (1979).
\bibitem{Freiman}
Y. A. Freiman,  A. Grechnev, S. M. Tretyak, A. F. Goncharov and R. J. Hemley, Phys. Rev. B {\bf 86}, 014111 (2012).
\bibitem{loubeyre}
P. Loubeyre, R. LeToullec, D. Hausermann, M. Hanfland, R. J. Hemley, H.-K. Mao and L. W. Finger, Nature {\bf 383}, 702 (1996).
\bibitem{aka}
Y. Akahama, M. Nishimura, H. Kawamura, N. Hirao, Y. Ohishi and K. Takemura, Phys. Rev. B {\bf 82}, 060101(R) (2010).
\bibitem{vinet}
P. Vinet, J. Ferrante, J. H. Rose and J. R. Smith, J. Geophys. Res. {\bf 92}, 9319 (1987).
\bibitem{lorenzana}
H. E. Lorenzana, I. F. Silvera and K. A. Goettel, Phys. Rev. Lett. {\bf 64}, 1939 (1990).
\bibitem{glyde}
See, for instance, R. T. Azuah, W. G. Stirling, H. R. Glyde, M. Boninsegni, P. E. Sokol and S. M. Bennington, Phys. Rev. B {\bf 56}, 14620 (1997).
\end{thebibliography}
\end{document}